\documentclass[12pt]{article}
\usepackage{graphics, setspace,cite}
\usepackage{mathrsfs}
\usepackage{amsfonts}
\usepackage{textcomp}
\usepackage{color}
\usepackage{amsmath}
\usepackage{graphicx}
\usepackage{amssymb}
\usepackage{authblk}

\usepackage{epstopdf}
\DeclareGraphicsExtensions{.eps}

\begin{document}
\thispagestyle{empty}

\def\theequation{\arabic{section}.\arabic{equation}}
\def\a{\alpha}
\def\b{\beta}
\def\g{\gamma}
\def\d{\delta}
\def\dd{\rm d}
\def\e{\epsilon}
\def\ve{\varepsilon}
\def\z{\zeta}
\def\B{\mbox{\bf B}}\def\cp{\mathbb {CP}^3}

\newcommand{\h}{\hspace{0.5cm}}

\begin{titlepage}

\renewcommand{\thefootnote}{\fnsymbol{footnote}}
\begin{center}
{\Large \bf Symmetry break in a scalar field theory in a
$(1 + 1)$-spacetime}
\end{center}
\vskip 1.2cm \centerline{\bf Gilberto N. S. Filho}

\vskip 10mm
\centerline{\sl Universidade Federal de Sergipe - UFS} 
\centerline{Centro de Ci\^encias Exatas e Tecnologia - CCET} 
\centerline{Departamento de F{\'i}sica}
\centerline{\sl Cidade Univ. Prof. Jos\'e Alo{\'i}sio de Campos} 
\centerline{\sl Av. Marechal Rondon, s/n, Jd. Rosa Elze}
\centerline{\sl S\~ao Crist\'ov\~ao/SE - Brazil.}
\vskip .5cm

\vspace*{0.6cm} \centerline{\tt gnsfilho@gmail.com}

\vskip 20mm

\baselineskip 18pt

\begin{center}
{\bf Abstract}
\end{center}

We consider in this work a relativistic scalar field theory in a $(1+1)$ Minkowski spacetime for a class of periodic potentials. These potentials exhibit solutions known as kinks and antikinks with topological charges, energy density, and mass dependent on the potential's parameters. We break the symmetry from a vacuum expectation value (\emph{vev}) zero for a non zero \emph{vev}, which means that the supersymmetry (SUSY) has spontaneously broken. 

\end{titlepage}
\newpage
\baselineskip 18pt

\def\nn{\nonumber}
\def\tr{{\rm tr}\,}
\def\p{\partial}
\newcommand{\non}{\nonumber}
\newcommand{\bea}{\begin{eqnarray}}
\newcommand{\eea}{\end{eqnarray}}
\newcommand{\bde}{{\bf e}}
\renewcommand{\thefootnote}{\fnsymbol{footnote}}
\newcommand{\be}{\begin{eqnarray}}
\newcommand{\ee}{\end{eqnarray}}

\vskip 0cm

\renewcommand{\thefootnote}{\arabic{footnote}}
\setcounter{footnote}{0}

\setcounter{equation}{0}
\section{Introduction}
The nonlinear phenomena in physics have attracted much interest with works in different kinds of potentials and they appear in different branches of physics as high energy physics and cosmology  \cite{Rajaraman,CS,TS,Belova,Bazeia1, Khoury,Yu, Paul,Hawking}.    These nonlinear  systems in $(1+1)$ dimension present topological solutions known as kinks (antikinks) solitons and are of general interest \cite{Bogdanov,Bazeia2,Bazeia2a,Bazeia3,Simas,Vakhid,Sordano,Cao,Mendonca,Faddeev}. We are considering a self-interacting relativistic real scalar field theory in a $(1+1)$ Poincaré-Minkowski spacetime for a class of trigonometric potentials that describes sine-Gordon (SG) and double sine-Gordon (DSG) models \cite{Bazeia2,Vakhid}.  These models appear very naturally within the context of the deformation procedure considered in \cite{Bazeia2a}. These potentials are dependent on a set of parameters that determines the solutions of the equations of motions. We determine the degenerated vacuum and the degenerated false vacuum of the potential in function of these parameters. The vacuum and false vacuum determine the topological sectors of the potential and we calculate the topological charge, energy density, and mass of these kinks for each topological sector.  We break the symmetry from degenerated vacuum expectation value (\emph{vev}) zero for a degenerated non zero \emph{vev} and determines the solutions for this class of potentials. The solution now became periodic with the kinks and the energy density no more spread in all space, they became compact \cite{compact1,Bazeia-lump,Bazeia-compact}. The kinks are now localized and we determine the charge and the mass of the kinks in a period. These solutions are Bogomol'nyi-Prasad-Somerfield (BPS) states \cite{BPS}, with the energy $E_{BPS}$ calculated in a period.  The symmetry break for a $vev \neq 0$ of the scalar potential is related to a spontaneous symmetry break of supersymmetry theories, with a superpotential related to the respective scalar potential \cite{Shifman-AQFT,Shifman-Yung-SUSY}. Before we describe the model we will in the next section review briefly the scalar field theory in a $(1+1)$ Minkowski spacetime. 

\section{Scalar Field Theory}

 We are working in a Minkowski $\mathbb{M}^2$ spacetime with $(1+1)$-dimension, with  Lorentzian flat metric $\eta_{\mu\nu}=\mbox{diag}(+,-)$, with $x^{\mu}=(t,\mathbf{x})$, $x^{2} = x_{\nu}x^{\nu}=\eta_{\mu\nu}x^{\mu}x^{\nu}=(x^{0})^{2}- (x^1)^{2}$.
 
  A scalar field theory with potential energy ${U}(\phi)$  is described by the Lagrangian density

\begin{equation}
{{\cal L}} = \frac{1}{2}\partial_\mu\phi\partial^\mu\phi-{U}(\phi).
\label{1a}
\end{equation}

For a Poicaré-Lorentz invariant  Lagrangian, we get the energy-momentum density tensor

\begin{equation}
T_{\mu\nu} = \partial_\mu\phi\partial_\nu\phi - \delta_{\mu\nu}{{\cal L}},
\end{equation}
\noindent that is conserved $\partial_\mu T^\mu_\nu=0$. From the energy-momentum tensor, we get the momenta

\begin{equation}
P^{\mu} = \int dx T^{\mu0},
\label{pi_mu}
\end{equation}
\noindent where $P^{0}$ is the energy of the system 
\begin{equation}
E = \int_{-\infty}^{+\infty}dx\left(\frac{1}{2}(\dot{\phi})^{2}+\frac{1}{2}(\phi')^{2}+U(\phi)\right),
\label{p0}
\end{equation}
\noindent and $P^{1}$ is the momentum 
\begin{equation}
P = - \int_{-\infty}^{+\infty}dx \dot{\phi}\phi'.
\label{p1}
\end{equation}

Using the Euler-Lagrange equation, we can get the equation of motion from Lagrangian (\ref{1a})
 
\begin{equation}
 \square\phi = -\frac{dU}{d\phi},
\label{2a}
\end{equation}
\noindent where $\square$ stand for the D'Alambertian symbol, with

\begin{equation}
 \square\phi = \ddot{\phi}-\phi''.
\label{3a}
\end{equation}

For a static solution, $\dot{\phi}=0$, we get from Eqs. (\ref{2a}) and (\ref{3a}) that

\begin{equation}
U(\phi) = \frac{1}{2}\left(\phi'\right)^{2}.
\label{5a}
\end{equation}

By integrating  the Eq. (\ref{5a}) we get the equation for the static solution

\begin{equation}
x-x_{0}=\pm\int_{\phi(x_{0})}^{\phi(x)}\frac{1}{\sqrt{2U(\tilde{\phi})}}d\tilde{\phi}.
\label{6a}
\end{equation}

In the following sections, we will solve the Eq. (\ref{6a}) for a model with trigonometric potential.

\section{The model}
\begin{figure}[ht]
\begin{center}
\includegraphics[scale=0.6,angle=0]{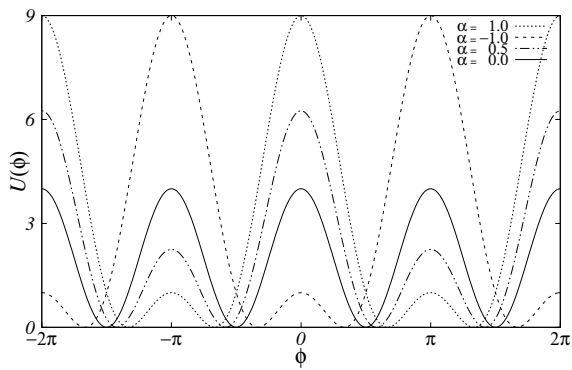}
\caption{The potential (\ref{pot}) for some values of the parameters $\alpha$ and $\beta=2$,  and $\omega =1$.}
\label{pots-bnbs}
\end{center}
\end{figure}

We will consider the following potential 

\begin{equation}
U(\phi) = \left(\alpha + \beta\cos(\omega\phi)\right)^{2},
\label{pot}
\end{equation}
\noindent where $\omega > 0$ is the frequency of the potential and the parameters $\alpha$ and $\beta$ determines the vacuums and the false vacuums of the potential. The potential (\ref{pot}) is positive  semi-definite, $U(\phi)\geq 0$, and for  $\beta^2 > \alpha^2$ it has infinitely degenerated vacuums for $\phi_{vac} = \frac{2n\pi}{\omega} + \frac{1}{\omega}\arccos(- \frac{\alpha}{\beta})$ and infinitely degenerated two false vacuums  for $\phi_{fvac} = \frac{2n\pi}{\omega}$ ($n\in \mathbb{Z}$) in each topological sector of the potential. There is a local false vacuum (\emph{lfvac}) $U_{lfvac} = (\beta - \alpha)^2$ for $n$ odd and a global false vacuum (\emph{gfvac})  $U_{gfvac} = (\beta + \alpha)^2$ for $n$ even. These two false vacuums determines the two topological sectors of the potential.   If $\alpha = 0$ the local false vacuums disappear  and we get the degenerated false vacuums $U_{fvac} = \beta^2$ in $\phi_{fvac} = \frac{2n\pi}{\omega}$  and the degenerated vacuums for $\phi_{vac} = \frac{(2 n + 1)\pi}{2\omega}$. If $\alpha = \beta$ we get also one degenerated false vacuum $U_{fvac} = 4\alpha^2$ in $\phi_{fvac} = \frac{2n\pi}{\omega}$  and degenerated vacuum for $\phi_{vac} = \frac{(4 n + 1)\pi}{\omega}$. For $\beta^2 > \alpha^2$, if we change the sign of $\alpha$ or $\beta$ we exchange the local vacuum and the global vacuum and so we exchange the two topological sectors of the potential. This is equivalent to change $\phi \rightarrow \phi \pm \frac{(2n + 1)\pi}{\omega}$.  The potential is invariant if we change the sign of both parameters $\alpha$ and $\beta$ (both negative) so, we will consider $\beta > 0$.  In the Fig. (\ref{pots-bnbs}) we show the potential (\ref{pot}) for some values of the parameters. 

The potential (\ref{pot}) is invariant under the $\mathbb{Z}_2$ symmetry $\phi \rightarrow - \phi$ and because its periodicity it is invariant under the translation $\phi \rightarrow  \phi + \frac{2n\pi}{\omega}$ and $n\in \mathbb{Z}$. These discrete transformations don't change the Hamiltonian density neither the Lagrangian density of the system and so the energy density is invariant under parity transformation and we get two solution,  $\phi_{\pm}$, for the equation of motion (\ref{2a}).

To get the static solution we need solve the equation (\ref{6a}) for the potential (\ref{pot}). The solutions are the topological solitons known as kinks, $\phi_{+}$, and antikinks, $\phi_{-}$. For $\alpha=\beta$ we get the solutions

\begin{equation}
\phi_{\pm}(x) = \pm \frac{2}{\omega}\arctan\left\{\sqrt{2} |\alpha| \omega x \right\} \pm \frac{2n\pi}{\omega}.
\label{kinks-sg}
\end{equation}

The energy density for  $\alpha=\beta$ is

\begin{equation}
\rho(x) = \frac{|\beta|^{2}}{\left(1 + 2\omega^{2}|\beta|^{2}x^{2}\right)^{2}},
\label{e-dens-1}
\end{equation}
\noindent and the mass of the solitons is

\begin{equation}
M = \frac{2\sqrt{2}\pi|\beta|}{\omega}.
\label{mass-1}
\end{equation}

\begin{figure}[ht]
\begin{center}
\begin{tabular}{cc}
\includegraphics[scale=0.6,angle=0]{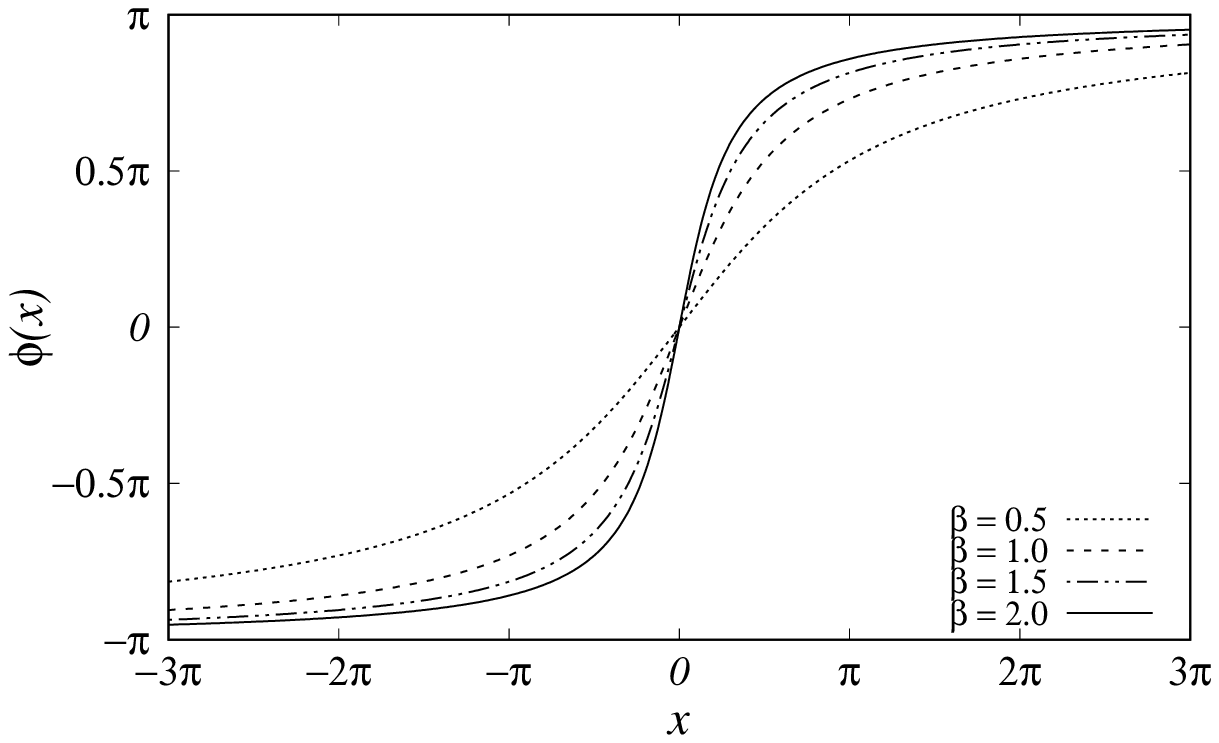} \\
\includegraphics[scale=0.6,angle=0]{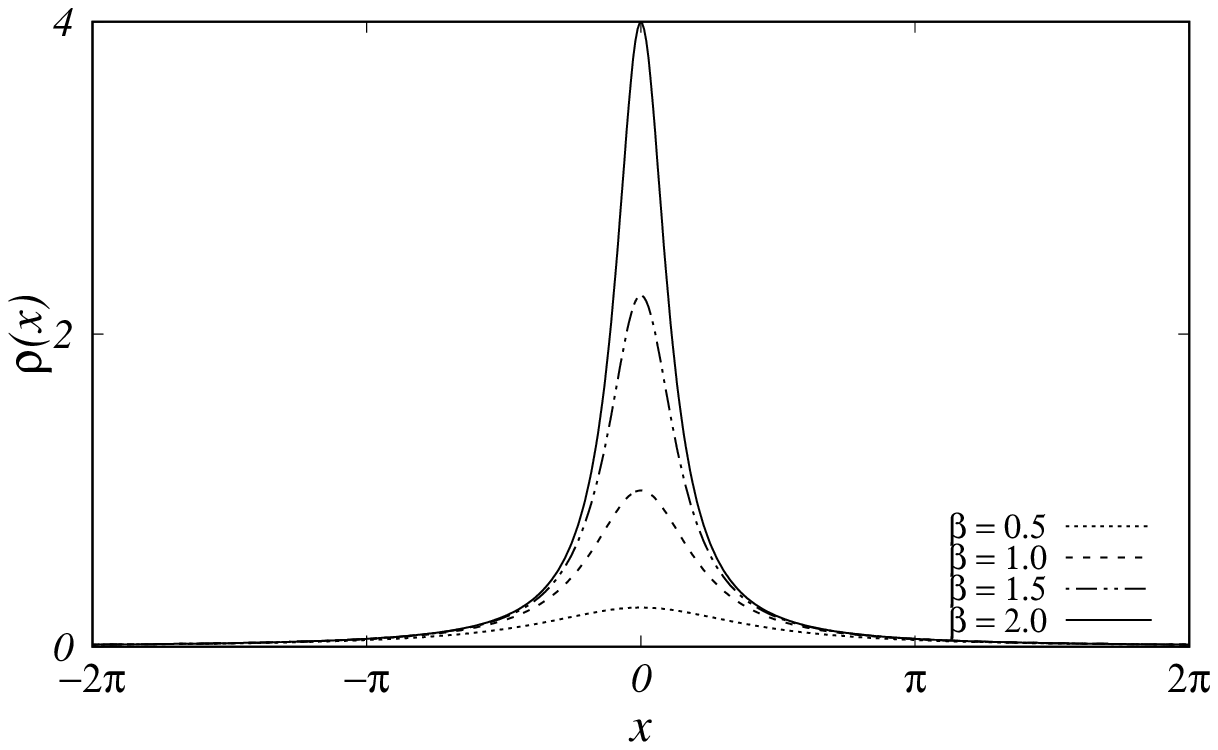}
\end{tabular}
\caption{The kinks solutions and energies densities for $\alpha = \beta$ and $\omega = 1$.}
\label{dens1a}
\end{center}
\end{figure}

These kinks and antikinks solutions have conserved topological charges, $Q_{\pm}$, which also implies that there is  a conserved topological current, $\partial_\mu J^\mu_T = 0$. This current is defined by

\begin{equation}
J^\mu_T = \frac{1}{2} \varepsilon^{\mu\nu} \partial_\nu \phi.
\end{equation}
\noindent  The topological charge is defined by

\begin{equation}
Q_{\pm} =  \phi_{\pm}(x\rightarrow +\infty) -  \phi_{\pm}(x\rightarrow -\infty).
\label{Q-def}
\end{equation}

Taking the limits $x\rightarrow \pm \infty$ in the Eq. (\ref{kinks-sg}), for $\alpha=\beta$,  we get the topological charges  $Q_{\pm} = \pm \frac{2\pi}{\omega}$. 

In Fig. (\ref{dens1a})  we show the kink solution and the energy density for some values of the parameters of the potential (\ref{pot}).

Using the potential (\ref{pot}) and the following equation
\begin{equation}
U(\phi)=\frac{1}{2}\left(\frac{d\mathit{W}}{d\phi}\right)^{2},
\label{sup-pot1}
\end{equation}
\noindent  we get the superpotential  

\begin{equation}
\mathit{W}=\pm\sqrt{2}\left[\alpha \phi+\frac{\beta}{\omega}sen(\omega\phi)\right].
\label{sup-pot2}
\end{equation}
\noindent The solutions of the first-order equation (\ref{sup-pot1}) are the BPS states \cite{BPS}. They are stable, minimum energy static solutions. We can write the energy for the static solution using the superpotential $\mathit{W}$

\begin{equation}
E(x) = E_{BPS} + \frac{1}{2} \int_{-\infty}^{+\infty} dx \left( \frac{d\phi}{dx} \mp \frac{d\mathit{W}}{d\phi} \right)^2,
\end{equation}
\noindent where 
\begin{equation}
E_{BPS} = \left|\mathit{W}\left(\phi(+\infty)\right) - \mathit{W}\left(\phi(-\infty)\right)\right|,
\label{BPS-def1}
\end{equation}
\noindent  is the bound in the energy which only depends on values of the kinks (antikinks) in the boundary, vacuum, of each topological sector. Using the Eq. (\ref{Q-def}) and the odd parity of the kink (antikink) solution, we also can write the Eq. (\ref{BPS-def1}) as a function of the topological charge  

\begin{equation}
E_{BPS} = \left|\mathit{W}\left(\frac{Q_{\pm}}{2}  \right) - \mathit{W}\left(-\frac{Q_{\pm}}{2}   \right)\right|.
\label{BPS-def2}
\end{equation}

  Using the superpotential (\ref{sup-pot2}), for $\alpha=\beta$,  we found $E_{BPS} = \sqrt{2}|\alpha Q_{\pm}|$.
  

\section{Breaking the symmetry}

\begin{figure}[ht]
\begin{center}
\begin{tabular}{cc}
\includegraphics[scale=0.6,angle=0]{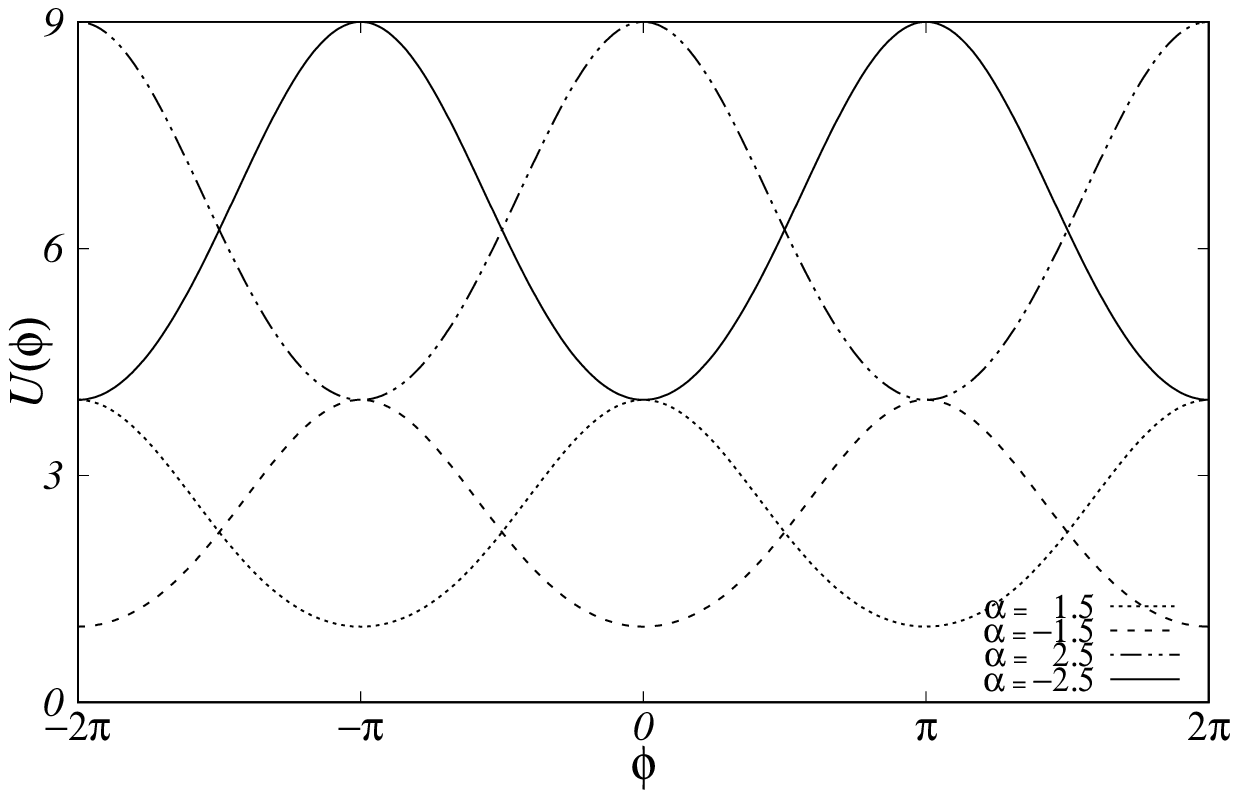} \\
\includegraphics[scale=0.6,angle=0]{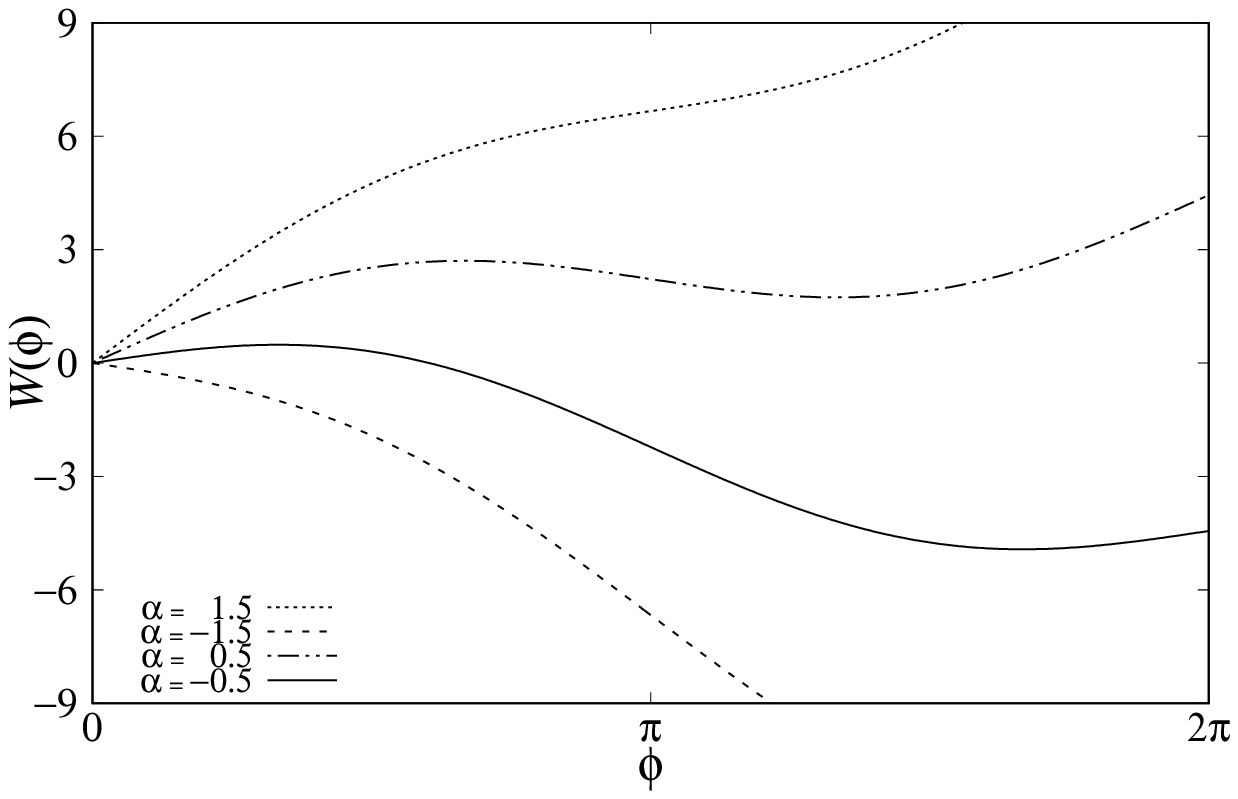}
\end{tabular}
\caption{The potential (\ref{pot}) and the superpotential (\ref{sup-pot2}) with symmetry break, for some values of the parameter $\alpha$, with $\beta=1$ and $\omega =1$.}
\label{pot1sb}
\end{center}
\end{figure}

\begin{figure}[ht]
\begin{center}
\begin{tabular}{cc}
\includegraphics[scale=0.6,angle=0]{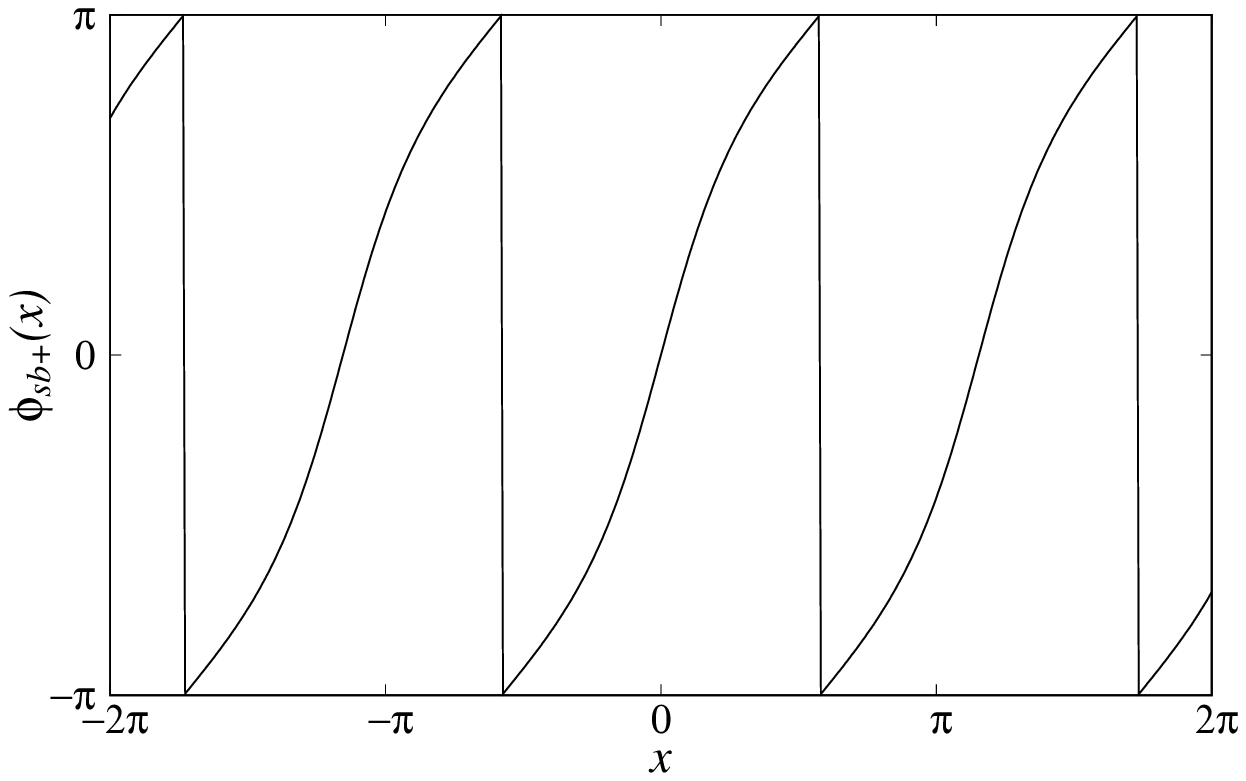} \\
\includegraphics[scale=0.6,angle=0]{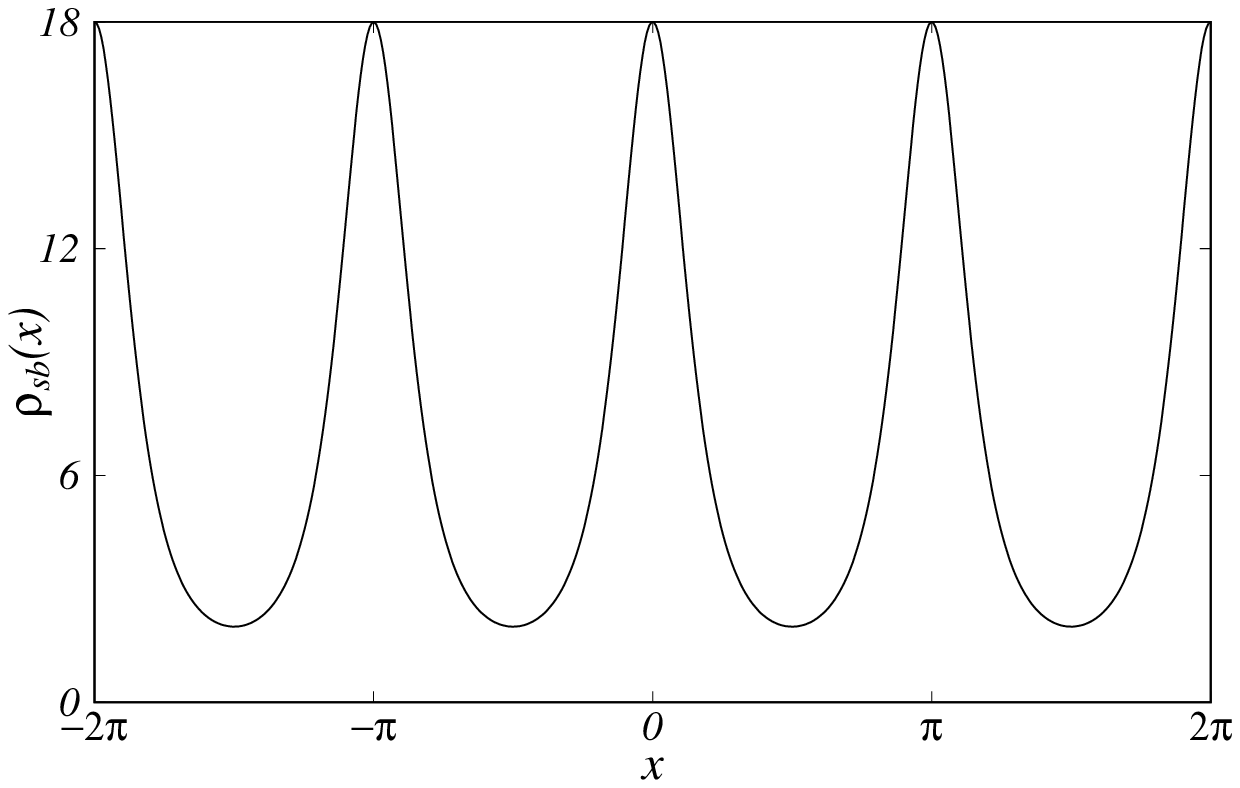}
\end{tabular}
\caption{The periodic kink and the energy density for the potential (\ref{pot}) with symmetry break, for  $\alpha = 2$, $\beta = 1$,  and $\omega =1$.}
\end{center}
\label{dens1c}
\end{figure}

Choosing $\alpha>\beta$ in the potential (\ref{pot}) we break the symmetry from a $vev=0$ to a $vev\neq0$  (Fig. \ref{pot1sb}). We get the vacuum value $U_{vac} = (\alpha - \beta )^2$ for $\phi_{vac} = \pm\frac{(2n+1)\pi}{\omega}$ and the false vacuum value $U_{fvac} = (\alpha + \beta )^2$ for $\phi_{fvac} = \pm\frac{2n\pi}{\omega}$. If we change the signal of $\alpha$ or $\beta$ we exchange the vacuum with the false vacuum. We get as solution  the periodic kinks (Fig. \ref{dens1c}), $\phi_{sb_{+}}$, and antikinks , $\phi_{sb_{-}}$,

\begin{equation}
\phi_{sb_{\pm}}(x)=\frac{2}{\omega}\mbox{tan}^{-1}\left\{ \mathcal{C}\;\mbox{tan}\left(\pm x s  \right)\right\} \pm \frac{2n\pi}{\omega}.
\label{sb-kak}
\end{equation}
\noindent with

\begin{equation}
\mathcal{C} \equiv \frac{\sqrt{\alpha^{2} - \beta^{2}}}{\alpha - \beta} \qquad\mbox{and}\qquad 
s = \sqrt{\frac{\omega^{2}}{2}\left(\alpha^{2} - \beta^{2}\right)}.
\end{equation}.

The topological charge is

\begin{equation}
Q_{\pm} = \phi_{sb_{\pm}}\left(x \rightarrow + \frac{\pi}{2s}\right) - \phi_{sb_{\pm}}\left(x \rightarrow - \frac{\pi}{2s}\right).
\label{Q-sb}
\end{equation}

Taking the limits $x\rightarrow \pm \frac{\pi}{2s}$ in the Eqs. (\ref{sb-kak}) we get the topological charges  $Q_{\pm} = \pm \frac{2\pi}{\omega}$.

We can write the energy for the static solution using the superpotential $\mathit{W}$

\begin{equation}
E(x) = E_{BPS} + \frac{1}{2} \int_{-\frac{\pi}{2s}}^{+\frac{\pi}{2s}} dx \left( \frac{d\phi}{dx} \mp \frac{d\mathit{W}}{d\phi} \right)^2,
\end{equation}
\noindent where 
\begin{equation}
E_{BPS} = 2\left|\mathit{W}\left(\frac{Q_{\pm}}{2}\right)\right|,
\end{equation}
\noindent is the bound in energy that only depends on the values of the kinks (antikinks) in the boundary, vacuum, of each topological sector.  Using the superpotential (\ref{sup-pot2}) we found $E_{BPS} = \sqrt{2}|\alpha Q_{\pm}|$ for the topological sector between $\phi = \pm \frac{\pi}{\omega}$, with $x = \pm \frac{\pi}{2s}$.

We can see in the Eqs. (\ref{sup-pot1}) and (\ref{sup-pot2}) that we have spontaneous supersimmetry break SSB if the potential (\ref{pot}) has $vev\neq 0$ \cite{Shifman-AQFT}. Therefore, if we break the symmetry of the potential (\ref{pot}) from a $vev=0$ to a $vev\neq 0$ we also have SSB.

The energy density became also periodic (Fig. \ref{dens1c})

\begin{equation}
\rho_{sb}(x) = 2\mathcal{D} \frac{\sec^{4}(\pm x\; s )}{\left\{ 1+\left[\mathcal{E}\;\mbox{tan}(\pm x \; s )\right]^{2}\right\}^{2}}.
\label{sb-ed}
\end{equation}
\noindent with

\begin{equation}
\mathcal{D} \equiv \frac{|\alpha^{2} - \beta^{2}|^{2}}{(\alpha - \beta)^{2}}, \qquad \mathcal{E} \equiv \frac{\sqrt{\alpha^{2} - \beta^{2}}}{\alpha - \beta},
\end{equation}
\noindent and the mass of the kink (antikink) is
\begin{equation}
M = \frac{\pi\sqrt{2}}{|\omega|}(\alpha - \beta)\left[\frac{|\alpha^{2} - \beta^{2}|}{(\alpha - \beta)^{2}}+1\right].
\label{sb-sm}
\end{equation}

\section{Summary}

We have started considering a relativistic self-interacting scalar field theory in a $(1+1)$ Minkowski spacetime for a class of periodic potentials with degenerated vacuums that break internal symmetries. We have broken the symmetry of the potential from a vacuum expectation value (\emph{vev}) zero for a non zero \emph{vev}, which means that supersymmetry (SUSY) has spontaneously broken. The solutions for this symmetry break are compact kinks (antikinks) with a periodic energy density. We determine the potential energies for the degenerated vacuums and false vacuums.

\section*{Acknowledgments}

The author acknowledges the Physics Department of the Universidade Federal de Sergipe. 


\end{document}